\documentclass[amsmath,preprint,endfloats]{revtex4-1}
\usepackage{graphicx}% Include figure files
\usepackage{dcolumn}% Align table columns on decimal point
\usepackage{hyperref}
\usepackage{units} %\nicefrac{a}{b}
\usepackage{epstopdf}
\hypersetup{
	colorlinks = true,
	urlcolor = blue
}
\begin{document}

%\preprint{}

\title{Hot carrier and hot phonon coupling during ultrafast relaxation of photoexcited electrons in graphene}

\author{J. M. Iglesias}
\affiliation{Department of Applied Physics, University of Salamanca, Salamanca 37008, Spain}
\author{M. J. Mart\'in}
\affiliation{Department of Applied Physics, University of Salamanca, Salamanca 37008, Spain}
\author{E. Pascual}
\affiliation{Department of Applied Physics, University of Salamanca, Salamanca 37008, Spain}
\author{R. Rengel}
\email[Electronic mail: ]{raulr@usal.es}
\thanks{Copyright 2016 AIP Publishing. This article may be downloaded for personal use only. Any other use requires prior permission of the author and AIP Publishing. The following article appeared in Appl. Phys. Lett. 108, 043105 (2016) and may be found at \url{http://dx.doi.org/10.1063/1.4940902}}
\affiliation{Department of Applied Physics, University of Salamanca, Salamanca 37008, Spain}

\date{\today}

\begin{abstract}
We study, by means of a Monte Carlo simulator, the hot phonon effect on the relaxation dynamics in photoexcited graphene and its quantitative impact as compared to considering an equilibrium phonon distribution. 
Our multi-particle approach indicates that neglecting the hot phonon effect significantly underestimates the relaxation times in photoexcited graphene. 
The hot phonon effect is more important for a higher energy of the excitation pulse and photocarrier densities between $1$ and $3\times 10^{12} \mathrm{~cm}^{-2}$. 
Acoustic intervalley phonons play a non-negligible role, and emitted phonons with wavelengths limited up by a maximum (determined by the carrier concentration) induce a slower carrier cooling rate. 
Intrinsic phonon heating is damped in graphene on a substrate due to additional cooling pathways, with the hot phonon effect showing a strong inverse dependence with the carrier density.
\end{abstract}

%\pacs{}% insert suggested PACS numbers in braces on next line

\maketitle %\maketitle must follow title, authors, abstract and \pacs

Graphene features, among other properties, large carrier mobility at room temperature along with gapless linear energy spectra for electrons and holes, that results in a linear optical absorption with virtually no photon wavelength restriction,\cite{Nair_Science_320_2008} making it a promising material for the development of a wide range of highly efficient photonic and optoelectronic applications, including those operating in the terahertz range. \cite{Ryzhii_JournalofAppliedPhysics_107_2010, Sensale-Rodriguez_NatureCommunications_3_2012, Cai_NatureNanotechnology_9_2014} 
%Consequently, an intense effort has been made in the recent years in order to get a good understanding of the carrier dynamics involved during and after photoexcitation.\cite{George_NanoLetters_8_2008, Sano_AppliedPhysicsExpress_4_2011, Breusing_PhysicalReviewB_83_2011, Lin_JournalofAppliedPhysics_113_2013, Winnerl_JournalofPhysics_CondensedMatter_25_2013, Tomadin_PhysicalReviewB_88_2013, Winzer_JournalofPhysics_CondensedMatter_25_2013} 
Consequently, an intense effort has been made in the recent years in order to get a good understanding of the carrier dynamics involved during and after photoexcitation\cite{George_NanoLetters_8_2008, Wang_AppliedPhysicsLetters_96_2010, Lin_JournalofAppliedPhysics_113_2013, Winnerl_JournalofPhysics_CondensedMatter_25_2013, Kim_PhysicalReviewB_84_2011, Tomadin_PhysicalReviewB_88_2013, Breusing_PhysicalReviewB_83_2011, Butscher_AppliedPhysicsLetters_91_2007, Winnerl_JournalofPhysics_CondensedMatter_25_2013, Sano_AppliedPhysicsExpress_4_2011,Satou2015} from a purely theoretical approach and also from an experimental point of view (pump--probe differential transmission spectroscopy) accompanied by means of various modelling techniques.
Right after photoexcitation, an ultrafast thermalization of the carriers takes place driven by Coulomb dual carrier scattering.\cite{Iglesias_IOPConferenceSeries_2015, Sano_AppliedPhysicsExpress_4_2011, Winzer_JournalofPhysics_CondensedMatter_25_2013} 
Simultaneosly, the carriers partially cool by transferring their energy to the graphene and substrate lattices by means of phonon cascade emissions.\cite{George_NanoLetters_8_2008, Breusing_PhysicalReviewB_83_2011, Winzer_JournalofPhysics_CondensedMatter_25_2013, Winnerl_JournalofPhysics_CondensedMatter_25_2013} 
As a result, a hot thermal distribution of electrons and holes is achieved in tens of $\mathrm{fs}$. \cite{ Tomadin_PhysicalReviewB_88_2013, Lin_JournalofAppliedPhysics_113_2013,  Winnerl_JournalofPhysics_CondensedMatter_25_2013} 
Later processes involve the cooling of the carriers as a consequence of scattering with phonons in the tail of the energy distribution, at the same time that recombination leads the system towards the full thermodynamic equilibrium.\cite{Breusing_PhysicalReviewB_83_2011, George_NanoLetters_8_2008}
Several authors remark the importance that the hot phonon (HP) effect would have on this dynamics. \cite{Wang_AppliedPhysicsLetters_96_2010, Lin_JournalofAppliedPhysics_113_2013, Huang_SurfaceScience_605_2011} 

The ensemble Monte Carlo (EMC) technique has been employed to study the influence of HP on the static transport properties of monolayer graphene under high--field conditions.\cite{Fang_PhysicalReviewB_84_2011} 
This method has been proven also to be worthy in the study of ultrafast carrier dynamics in a sub--ps time scale in other materials.
%%%
For example, in GaAs this dynamics is examined in a femtosecond scale by Zhow\cite{Zhou1990}, while the effects of carrier screening are disentangled by Osman\cite{Osman1987} and the non stationary dynamics in THz sources are investigated by Kim\cite{KimPhD} with an EMC simulator coupled to a Poisson solver. Also, in more complex structures such as superlattices\cite{Rossi1995}, this technique has been employed to describe the carrier relaxation and transport. Parallel phonon and electron EMC simulations are used by Tea\cite{Tea2011} to examine the relaxation of carriers and phonons in highly excited polar semiconductors. The transient characteristics of carbon nanotubes up to a few picoseconds are investigated by Verma.\cite{Verma2005}
%%%
As it concerns graphene, ultrafast relaxation phenomena have been investigated within a semiclassical framework by means of the Boltzmann transport equation; for example, Kim\cite{Kim_PhysicalReviewB_84_2011} and Low\cite{Low_PhysicalReviewB_2012} were able to predict the dynamics during and after photoexcitation, and 
%The response of the material to a high electric THz pulse has been investigated too [19]. 
Satou\cite{Satou2015} examined the effects of carrier--carrier scattering on photoinverted carriers in graphene for different substrates. 
On its behalf, the EMC technique has been employed by Sano \cite{Sano_AppliedPhysicsExpress_4_2011} to give insight into the carrier--carrier and carrier--phonon processes in photoexcited graphene.

Nonequilibrium phonons have been previously accounted for in the study of photocarrier relaxation dynamics\cite{Butscher_AppliedPhysicsLetters_91_2007,Winnerl_JournalofPhysics_CondensedMatter_25_2013}.
However, there is not yet a detailed analysis from a microscopic point of view, about their impact during this process  and which are the quantitative differences found with regard to considering a phonon distribution in equilibrium used in other studies. \cite{Kim_PhysicalReviewB_84_2011,Sano_AppliedPhysicsExpress_4_2011,Low_PhysicalReviewB_2012} 
In this work, we address the HP effect during the thermalization and cooling stages of the carrier relaxation dynamics, evaluating its impact for  different photocarrier concentrations and photon wavelengths.
With this purpose the time interval is considered up to $2\mathrm{~ps}$ after photoexcitation, under the assumption that the carrier recombination is negligible during this interval. \cite{George_NanoLetters_8_2008} 

The pertinent interactions included in our EMC model\cite{Rengel_AppliedPhysicsLetters_104_2014, Rengel_JournalofAppliedPhysics_114_2013} are the acoustic intravalley, optical and acoustic intervalley phonons. 
The scattering rates for the TO/LO and TA/LA phonon modes are calculated within an electron plane wave model within the deformation potential approximation\cite{Hwang2008}, which is a standard approach to determine the scattering rates for numerical models in graphene.
The physical parameters are adjusted in order to fit the rates calculated from the first--principles density functional theory.\cite{Sule2012,Borysenko_PhysicalReviewB_2010} 
It is known that the Born--Oppenheimer approximation breaks down in graphene\cite{Pisana2007}, however in the case of photoexcited graphene the average energy of the electron system is typically much larger than the Fermi energy in equilibrium during most of the relaxation timespan, so we do not expect this approximation to be a severe restriction for the approach presented here. 
In addition, due to the strong coupling between carriers in graphene, short range carrier--carrier scattering is taken into account through a static screened Coulomb potential model. \cite{Li_AppliedPhysicsLetters_97_2010} 
In the case of graphene on a substrate, the additional surface polar phonon (SPP) modes, which are also responsible for the carrier relaxation, are also included. \cite{Low_PhysicalReviewB_2012} 
Screening is accounted for in the Random Phase Approximation\cite{Hwang_PhysicalReviewB_75_2007} by numerically calculating the Lindhard polarizability in its static limit from the energy distribution every time step in the EMC simulation. 
Since the most relevant effect of considering dynamic screening is the suppression of Auger processes\cite{Tomadin_PhysicalReviewB_88_2013}, which are forbidden in our model, we consider that the use of static screening should not severely compromise the study of the hot carrier and phonon coupling. 
To examine the hot--phonon effect arising from the coupled carrier--phonon dynamics, the out--of--equilibrium populations of the considered phonons are treated dynamically by accounting for the electron assisted emissions and absorptions resulting from the corresponding scattering mechanisms that take place during each time step. 
The spontaneous phonon decay is considered within the relaxation time approximation,\cite{Fang_PhysicalReviewB_84_2011} taking constant mean phonon lifetimes of $\tau_\mathrm{OP} = 2.5\mathrm{~ps}$ and $\tau_\mathrm{AC} = 5\mathrm{~ps}$, and $\tau_\mathrm{SPP} = 5\mathrm{ ps}$.\cite{Wang_AppliedPhysicsLetters_96_2010, Bonini_NanoLetters_12_2012, Lindsay_PhysicalReviewB_89_2014, Katti_JournalOfAppliedPhysics_2013}
The scattering probabilities are recalculated every time step, which is set to $0.5\mathrm{~fs}$, considering the updated phonon population and dielectric function. 
The number of simulated particles ranges from $50000$ to $1000000$ depending on the carrier density. 

Various pumping photon wavelengths and carrier concentrations (which would be the result of different pump fluences) are considered, setting an initial electron distribution according to the expression $
f(\varepsilon) = \mathcal{F}(\varepsilon, T, \mu)
+ \varphi_\mathrm{max} \exp \left [ 
	-(\varepsilon - \nicefrac{1}{2\,}\varepsilon_\mathrm{photon} )^2/2 \sigma^2
\right ]
$, \cite{Tomadin_PhysicalReviewB_88_2013}
where the first member of the right hand side is the Fermi equilibrium distribution for undoped graphene ($\mu = 0$) at $T=300\mathrm{~K}$, and the second represents the excited carriers arising from optical pumping of a light source with an energetic spectra peaked at $\varepsilon_\mathrm{photon}$, and where $\sigma=0.1\mathrm{~eV}$ accounts for the spectral bandwith of a Fourier limited pulse with a minimum temporal full width at half maximum of around $8\mathrm{~fs}$. The parameter $\varphi_\mathrm{max}$ is the maximum occupation at $\varepsilon= \nicefrac{1}{2\,}\varepsilon_\mathrm{photon}$ and relates the pumped concentration with the radiation fluence and the pulse energy. 

\begin{figure}[ht!]
	\centering
	\includegraphics{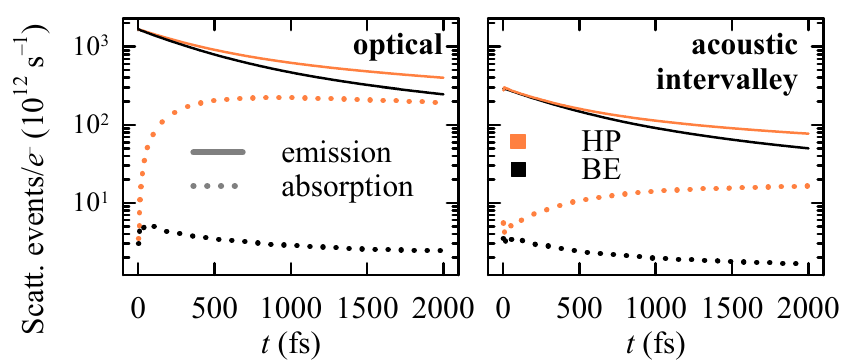}
	\caption{Temporal record of emission (solid line) and absorption (dotted line) scattering events considering electron and phonon population coupling (HP, orange lines) and the equilibrium Bose--Einstein distribution (BE, black lines) for $n_s=2\times 10^{12}\mathrm{~cm^{-2}}$ and $\varepsilon_\mathrm{pump}=1.2\mathrm{~eV}$ with the optical mode (left) and acoustic intervalley (right) phonon modes. 
		\label{fig:2}
	}
\end{figure}

When carriers are in an out--of--equilibrium situation, as it is the case of photoinversion, they lose energy by emitting phonons at a faster rate than they decay into other phonon modes, thus driving their population out--of--equilibrium as well.\cite{Wang_AppliedPhysicsLetters_96_2010, Huang_SurfaceScience_605_2011} 
The origin of the HP effect in one of the cases under study is shown in {figure~\ref{fig:2}}. 
Tracking of the electron scattering events with phonons reveals that phonon absorption for the optical mode is two orders of magnitude higher for almost all the time span while no such a huge difference is found for the emissions. 
For the acoustic intervalley mode, the phonon absorption scattering rate discrepancy between the HP and BE models is not so severe as in the optical mode due to its lower energy and coupling with hot carriers.
As an overall consequence, the net power dissipated by scattering with the lattice vibrations is reduced in comparison with the case in which the phonons are in equilibrium. 

\begin{figure}[h!]
	\centering
	\includegraphics{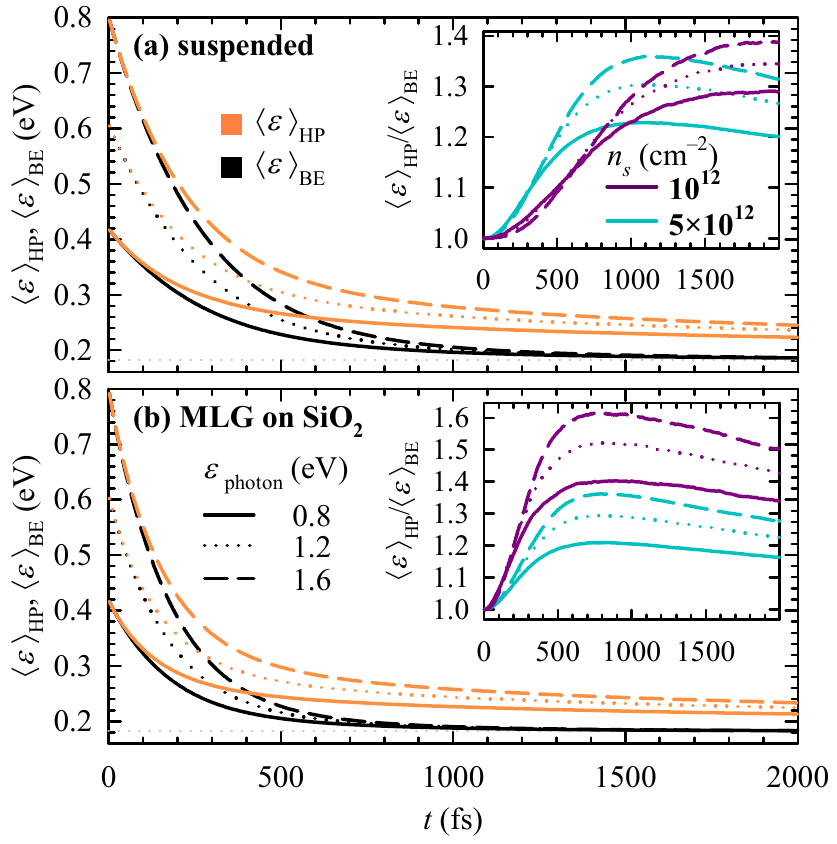}
	\caption{Temporal evolution of the ensemble average energy for a photoexcited carrier concentration equal to $10^{12}\mathrm{~cm^{-2}}$, in (a) suspended graphene and (b) graphene on SiO$_2$  considering carrier and HP coupling, $\langle \varepsilon \rangle_\mathrm{HP}$ (orange lines) and an equilibrium Bose--Einstein phonon distribution, $\langle \varepsilon \rangle_\mathrm{BE}$ (black lines) for comparison. 
		The insets show the ratio of these two quantities to help readers characterize the pulse wavelenght and photocarrier concentration dependence. 
		\label{fig:1}
	}
\end{figure}
To describe the temporal evolution of the ensemble, {figure~\ref{fig:1}} represents the average electron energy considering both a \mbox{hot carrier--hot phonon} (HP) coupled system, $\langle \varepsilon \rangle_\mathrm{HP}$, together with the average energy obtained with a Bose--Einstein (BE) equilibrium phonon occupation, $\langle \varepsilon \rangle_\mathrm{BE}$, for comparison. 
The timespan of this process is limited up to the instant when the average energy reaches the value corresponding to the thermal distribution with the quasi Fermi level established by the pumped concentration. 
An energy decay that tends towards a minimum thermal value is observed for all the cases under study. 
If the HP effect is included, the energy decay rate is slower as compared with the BE case, standing out the importance of this coupling, that prevents a faster relaxation. 
The insets shows the ratio between the two quantities in order to discern the impact of the HP effect with the pulse energy and the photocarrier density. 
In all cases under study, it is observed that the HP effect tends to increase as the photon energy is larger. 
This is due to the fact that reaching a fixed quasi Fermi level from an excitation state resulting from a larger photon wavelength implies a larger number of phonon emissions per electron. 
In suspended graphene, the effect tends to be higher for concentrations between $1$ and $3\times10^{12}\mathrm{~cm^{-2}}$ (not shown in the graphs) reaching its maximum for longer times than with higher concentrations.
In the case of graphene on SiO$_2$ the energy ratio shows similar temporal trends for all cases, but a stronger inverse dependence with the photocarrier density.
This is explained through the scattering probability of the intrinsic and SPP phonon modes, as for the firsts, the scattering probability scales linearly with the carrier energy, while for the latter it rises quickly at low energies and saturates for larger ones. 

Time--tracking of the phonon occupation is shown in figure 3. 
\begin{figure}[t!]
	\centering
	\includegraphics{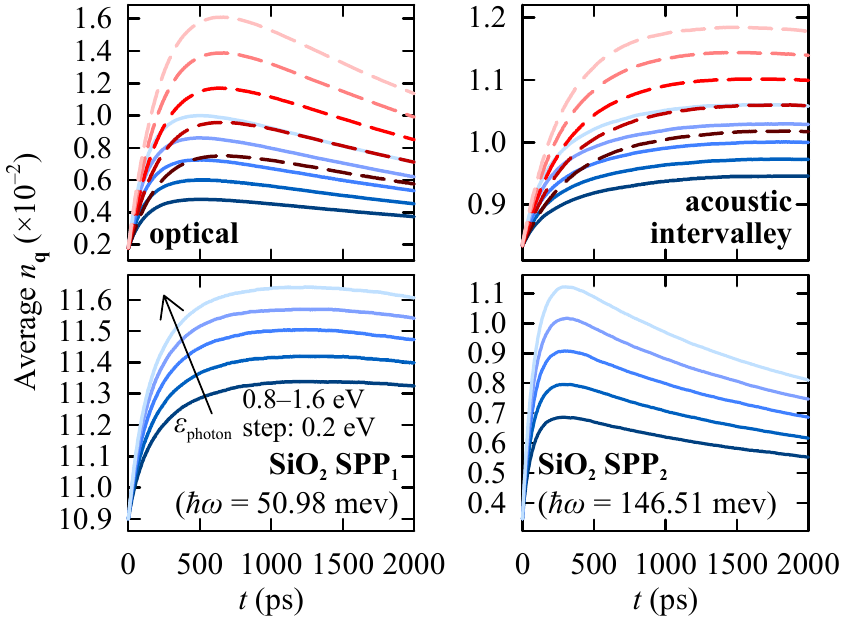}
	\caption{Averaged phonon occupation in the FBZ as a function of time after photoexcitation for (red--dashed lines) suspended graphene and (blue--solid lines) graphene on SiO$_2$ with a photocarrier density of $10^{12}\mathrm{~cm}^{-2}$.
	\label{fig:3}
	}
\end{figure}
The optical mode shows a fast growth during the first $500\mathrm{~fs}$ and then it drops linearly.
The acoustic intervalley mode manifest a less abrupt behaviour; the initial rise slope is less pronounced and in the time frame considered there is no net reduction, although it tends to saturate around $t=2.0\mathrm{~ps}$. 
When the monolayer graphene is supported on SiO$_2$ it can be seen that the heating of the intrinsic modes is more moderate. 
With regard to the substrate polar phonons, the SPP$_1$ mode shows a trend similar to that of the intrinsic acoustic intervalley, while the SPP$_2$ mode resembles the trend of the intrinsic optical one. 
The origin of these differences are, in the first place, the different electron-phonon couplings.
These are, for the intrinsic modes, a consequence of the distinct deformation potentials, making the intrinsic optical mode dominate over the acoustic intervalley. 
With regards to the SPPs, electron--phonon coupling is proportional to the phonon energy. 
Secondly, the larger phonon occupation in equilibrium for the less energetic ---i.e., the acoustic and SPP$_2$--- modes favours a bigger absorption/emission proportion, and thus a slower and smoother heating.
Finally, the shorter lifetime of the optical mode drives the population towards equilibrium through phonon decay faster than in the acoustic case. 
\begin{figure}[t!]
	\centering
	\includegraphics{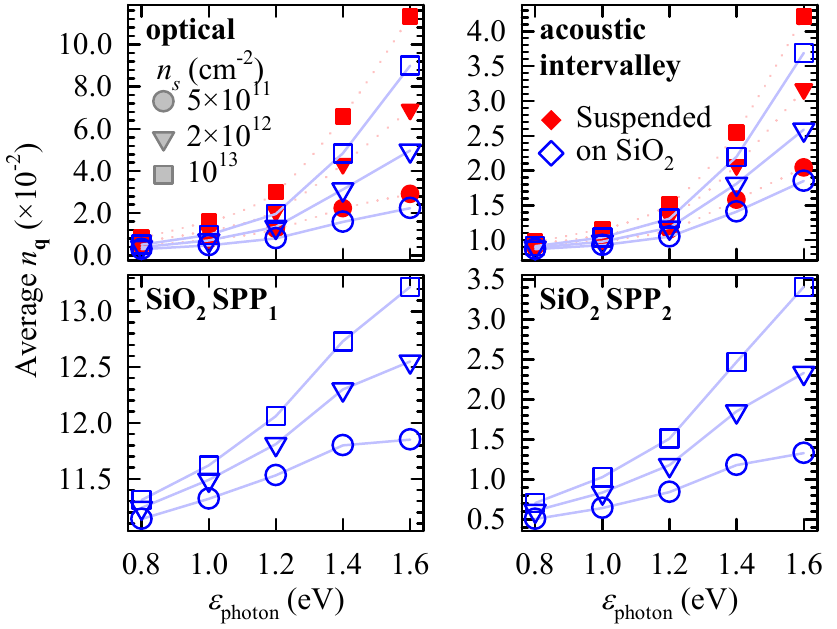}
	\caption{Averaged occupation in the FBZ at $t=750\mathrm{~fs}$ as a function of the photon energy for photocarrier concentrations of $5\times10^{11}$, $2\times10^{12}$ and $10^{13}$ $\mathrm{cm}^{-2}$.
	Red symbols stand for suspended graphene, and blue ones for graphene on SiO$_2$.
		\label{fig:4}
	}
\end{figure}
The evolution of the phonon population with the pumping pulse wavelength and electron concentration supports the statements about the energetic evolution of the electron ensemble discussed in figure~\ref{fig:1}. 
Phonon population dependences with the photocarrier density and the pulse energy can be seen in figure 4.
Reduced intrinsic optical and acoustic heating phonon for graphene on SiO$_2$ as a consequence of the additional cooling pathways is observed. 
The intrinsic optical mode is the most displaced from equilibrium, reaching mean phonon occupations 60 times larger than in equilibrium, followed by the SPP$_2$ mode ($10\times$), the acoustic intervalley ($5\times$) and finally, the SPP$_1$ mode, which barely heats ($1.2\times$).

\begin{figure}[t!]
	\centering
	\includegraphics{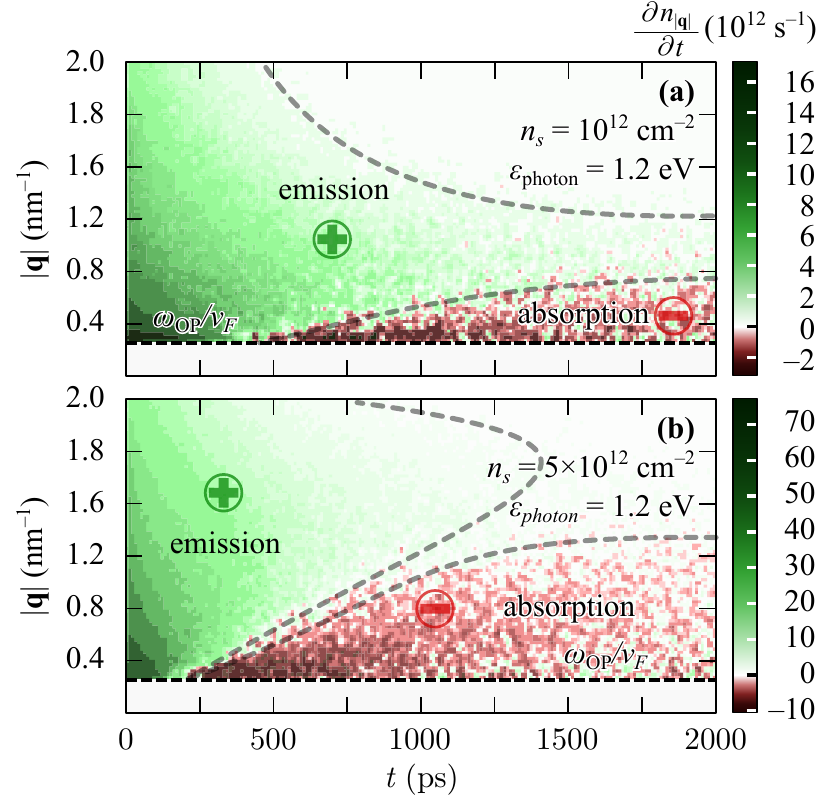}
	\caption{Phonon creation rate as a function of the time after photoexcitation with $\varepsilon_\mathrm{photon}=1.2\mathrm{~eV}$ and $|\mathbf{q}|$ for $n_s=$ (a) $10^{12}\mathrm{~cm^{-2}}$ and (b) $5\times10^{12}\mathrm{~cm}^{-2}$. Green areas stand for phonon creation, while red areas for their absorption.
	Dashed lines indicate approximately the limits of each region. 
	\label{fig:5}
	}
\end{figure}

Let us now focus on the phonon momentum resolved dynamics as shown in {figure~\ref{fig:5}}.
In the graphs, only the evolution of the phonons as a consequence of scattering with electrons is shown, although their natural decay within the corresponding lifetimes is considered as well in the model. 
At low concentrations (figure~\ref{fig:5}--a), phonons start being emitted with the highest rate during the first instants with a wavevector distribution focused in the smallest possible required by the phonon energy $|\mathbf{q}| \geq \omega_{\mathrm{OP}}/v_F$. 
At around $600\mathrm{~fs}$ the electron distribution is close to equilibrium, and phonon emission is mostly forbidden by the Pauli excussion principle, preventing energetic carriers to occupy already nearly full states close to the Dirac point. 
Instead, these electrons are more likely to absorb the previously emitted phonons, which gives rise to the net absorption rate with short wavevectors. This is reflected in the area with a net destructive phonon rate that starts to appear at $500\mathrm{~fs}$ and is constrained to $|\mathbf{q}| \leq 0.8\mathrm{~nm}^{-1}$.
In the high concentration case (figure~\ref{fig:5}--b), the phonon emission stage is much briefer than in the previous case, ending at around $200\mathrm{~fs}$ due to the high degeneracy of the system. 
Then, phonons start to be absorbed with a wavevector extent that grows with time, but reaches a limit at around $|\mathbf{q}| \leq 1.3\mathrm{~nm}^{-1}$. 
In both cases, it can be seen that phonons with short wavelengths do not seem to be ever absorbed, and in fact they are not, as such transitions are delimited by the energy of the carriers at late stages of the process. 
Instead, they decay naturally, and only play a role in the early stages, absorbing the energy of the electrons.
Regarding the pumping wavelength only quantitative differences can be noted in emission/absorption scales and in the range of the $|\mathbf{q}|$ involved, which is longer the shorter is the pumping wavelength. 
It has to be pointed out that phonons with wavevectors shorter than $\hbar \omega / v_F$ show no changes along the sampled time because interband processes are ruled out in the simulations. 
However, HPs are expected to play a role in slowing down the recombination processes as well. 

To summarize, we have studied the incidence of the transient dynamics of out--of--equilibrium phonons and charge carriers in graphene during the thermalization and intraband cooling stages of the relaxation process after photoexcitation. 
The HP effect manifests itself by reducing the energy transfer ratio from the carriers to the lattice. 
Its consequences grow with the photon energy and for moderate photocarrier densities. 
Phonon dynamics were also examined; the population of the intrinsic optical mode phonons reaches its maximum around 500 fs and decays thereafter, while the acoustic intervalley mode exhibits much slower dynamics. Phonons with large wavevectors emitted right after photoexcitation are difficultly reabsorbed when the thermal distribution is achieved by the electrons, while those with long wavelengths account for the complete intraband cooling bottleneck. When graphene lies on a substrate, intrinsic phonon heating becomes less severe and the HP effect clearly increases inversely with the carrier concentration due to the accompanying heating of the SPPs.
%The results shown indicate that considering the hot phonon effect is imperative in order to correctly describe the intraband relaxation process in photoexcited graphene, as well as taking into account the inelastic intervalley scattering interactions to capture the complete electron--phonon coupling of intrinsic graphene.

\begin{acknowledgments}% If you have acknowledgments, this puts in the proper section head
This work was supported by research project \mbox{TEC2013--42622--R} from the {Ministerio de Economía y Competitividad}.
\end{acknowledgments}

\printtables
\printfigures

\end{document}